\documentclass[prl,twocolumn,groupedaddress,floatfix]{revtex4}
\usepackage{amsmath}
\usepackage{graphicx}
\usepackage{bm}

\newcommand{\imag}{{\rm Im}}

\newcommand{\kb}{k_{\rm B}}

\newcommand{\bq}{{\bf q}}

\newcommand{\be}{\begin{equation}}
\newcommand{\ee}{\end{equation}}
\newcommand{\bea}{\begin{eqnarray}}
\newcommand{\eea}{\end{eqnarray}}
\newcommand{\bse}{\begin{subequations}}
\newcommand{\ese}{\end{subequations}}

\input{epsf}
\begin{document}

\title{Quantum critical scaling in graphene}
\author{Daniel E.~Sheehy}
\altaffiliation[Present address: ]{Department of Physics and Astronomy, Louisiana State University, Baton Rouge, LA 70803}
\author{J\"org Schmalian}
\affiliation{Ames Lab and Department of Physics and Astronomy, 
Iowa State University,
Ames, IA 50011}
\date{July 19, 2007}
\begin{abstract}
We show that the emergent relativistic symmetry of electrons in graphene
near its quantum critical point (QCP) implies a crucial importance of 
the Coulomb interaction.  We derive scaling laws, valid near the 
QCP, that dictate the nontrivial magnetic and charge response of 
interacting graphene.  Our analysis yields numerous 
predictions for how the Coulomb interaction will be manifested in
experimental observables such as  the diamagnetic response and 
electronic compressibility.
\end{abstract}

\maketitle

Recent experimental developments~have made possible the study of  graphene,
a single-atom thick sheet of graphite\cite{Novoselov04}. The novel
electronic properties of graphene arise from the linear, cone-shaped
energy-momentum dispersion of electrons at low
energies. This condensed
matter realization of a relativistic 
Dirac spectrum follows from simple models of
electrons hopping on the honeycomb lattice of graphene\cite{Semenoff84}, and
has been confirmed by a range of 
experiments~\cite{Novoselov05,Zhang05,Ohta07}.  

The relevant Hamiltonian is that of relativistic Coulomb-interacting fermions in two dimensions:
\begin{equation}
H=\sum_{l}v\widehat{\mathbf{p}}_{l}\mathbf{\cdot \sigma +}\frac{1}{2}
\sum\limits_{l\neq l^{\prime }}\frac{e^{2}}{\varepsilon \left\vert \mathbf{r
}_{l}-\mathbf{r}_{l^{\prime }}\right\vert },  \label{Ham}
\end{equation}
with velocity $v\simeq 10^{8}\mathrm{cm/}$\textrm{s}\cite{Novoselov05}. $%
\widehat{\mathbf{p}}_{l}=-i\hbar \nabla _{\mathbf{r}_{l}}$ is the momentum
operator and $\mathbf{\sigma }=\left( \sigma _{x},\sigma _{y}\right) $ are
 Pauli matrices that act in the space of the two sub-lattices of the
honeycomb lattice structure. There is an additional $N=4$ fold degeneracy
caused by spin and the two distinct nodes of the dispersion, with
the only tunable parameter in Eq.~(\ref{Ham}) being
 the dielectric constant $\varepsilon $.

 In case of the usual
electron gas, the first term in Eq.~(\ref{Ham}) is $\sum_{l}\widehat{
\mathbf{p}}_{l}^{2}/(2m)$. Then, dimensional arguments imply that the kinetic
energy dominates for high electron density while the Coulomb interaction
dominates at low density. The linear Dirac spectrum changes this situation.
The relative importance of the potential and kinetic energy is the same for
all densities and controlled by the dimensionless number $
\lambda =e^{2}/(4\varepsilon v\hbar )$. For $\lambda \ll 1$, the Coulomb
interaction is negligible. Using the above value for the electron
velocity yields $\lambda \simeq 0.55/\varepsilon $, i.e. $\lambda \simeq
0.55$ for a free standing graphene film in vacuum, implying that one cannot
ignore the Coulomb interaction. The role of interactions in graphene has
been discussed previously~\cite{Gonzalez94,Gorbar,Kvech06,Son07,DasSarma,Vafek07,Mishchenko,Barlas,Biswas}. 
However, few specific predictions for observable quantities have been made that allow for
a comparison with experiment (see, however, Ref.~\cite{Hwang0703}).
Here, we exploit the enlarged symmetry near its quantum critical point (QCP)
to deduce numerous predictions 
(based on scaling theory) of interacting graphene.

\begin{figure}%
\epsfxsize=8.5cm
\vskip.3cm
\centerline{\epsfbox{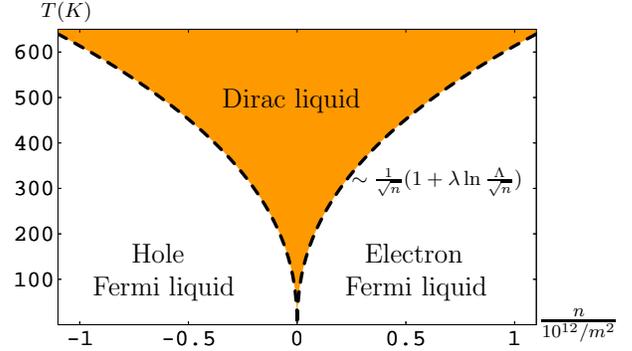}}
\vskip-.75cm
\caption{(Color Online) Quantum critical phase diagram of graphene as a function of density $n$ (in units of $10^{12} m^{-2}$)
and temperature $T$ (in $K$), for the vacuum case $\epsilon = 1$, 
showing the Dirac liquid and Fermi liquid
regimes separated by the crossover temperature $T^{\ast}$ [dashed lines, Eq.~(\ref{eq:tast})], 
with the quantum critical point occuring at $n = T = 0$.
}
\label{phasediagram}
\vskip-.45cm
\end{figure}

In this letter, we use a renormalization group (RG) approach to the Hamiltonian,
Eq.~(\ref{Ham}), and analyze the magnetic and charge response of graphene as a 
function of temperature $T$, carrier density $n$, chemical potential $\mu $
and magnetic field $B$. We make specific predictions for the 
 the compressibility $\kappa =\partial n/\partial\mu$, 
the diamagnetic susceptibility $\chi _{D}$, the magnetic moment $M(B)$, the heat capacity $C$, the
infrared conductivity $\sigma(\omega)$ and the
density-density correlation functions $\chi _{c}( \mathbf{q,}\omega)$. 
We demonstrate that interaction effects in these quantities are
measurable, allowing experiments to reveal the subtle interplay of
interactions and kinetic energy in a Dirac liquid. Our analysis is based on
the fact that for $T=B=\mu =n=0$, clean graphene is located at a QCP, as illustrated in Fig.~\ref{phasediagram},
 and its properties nearby can be obtained via crossover scaling arguments.

The low energy action that follows from Eq.~(\ref{Ham}) is 
\begin{equation}
\mathcal{S}=\hbar \int_{x}\psi ^{\dagger }\left( \partial _{\tau }\sigma
_{0}-iv\nabla _{\mathbf{r}}\mathbf{\cdot \sigma }\right) \psi +\frac{e^{2}}{%
2\varepsilon }\int_{x,x^{\prime }}\frac{n_{x}n_{x^{\prime }}}{\left\vert 
\mathbf{r-r}^{\prime }\right\vert }.  \label{action}
\end{equation}%
Here, $\psi =\psi \left( x\right) $ is a two component electron field where $
x=\left( \mathbf{r,}\tau ,s\right) $ stands for the $2D$ position $\mathbf{r}
$, imaginary time $\tau $, and valley and spin quantum numbers $s=1\cdots 4$%
, such that $\int_{x}\ldots =\int d^{2}\mathbf{r}\int_{0}^{\beta }d\tau
\sum_{s=1}^{4}\ldots $ with $\beta ^{-1}=\kb T/\hbar $. $n_{x}=\psi
^{\dagger }\left( x\right) \psi \left( x\right) $ is the electron density.

We perform a one-loop Kadanoff-Wilson RG analysis of 
Eq.~(\ref{action}). Fourier transforming $\psi(x) $ yields $\psi(k)$ where $k=(\mathbf{k,}\omega _{n},s)$ with
planar wave vector $\mathbf{k}$ and Matsubara frequency $\omega _{n}=(2n+1)\kb T/\hbar $. We trace out high energy modes with $\Lambda
/b<\left\vert \mathbf{k}\right\vert <\Lambda $ and obtain a renormalized
action. Due to the static nature of the Coulomb interaction, the one loop
fermion self energy $\Sigma \left( k\right) $ is frequency
independent, i.e. the fermion dynamics remains unrenormalized: $\partial
_{\tau }\rightarrow \partial _{\tau }$. Higher order diagrams\cite%
{Gonzalez94,DasSarma} or strong coupling effects\cite{Son07} cause $\Sigma
\left( k\right) $ to be $\omega $-dependent but are beyond the leading
diverging 
terms considered here. Similarly, vertex corrections describing
interactions between electrons and collective charge fluctuations vanish.
This is consistent with the Ward identity that follows from the conservation
of the total charge and implies $e^{2}\rightarrow e^{2}$ under
renormalization. The only nontrivial renormalization is that of the 
velocity. Since $\partial \Sigma \left( \mathbf{k},\omega =0\right) /\partial 
\mathbf{k}$ diverges logarithmically as $\left\vert \mathbf{k}\right\vert
\rightarrow 0$, within the RG approach this yields $
v\rightarrow v\left( 1+\lambda \log b\right) $. To complete the
RG procedure, we rescale the fermion field 
\begin{equation}
\psi (\mathbf{p},\omega ,s)\rightarrow Z_{\psi }\left( b\right) \psi (b%
\mathbf{p},Z_{T}^{-1}\left( b\right) \omega ,s)  \label{Zpsi},
\end{equation}
with $Z_{T}\left( b\right) =b^{-1}\left( 1+\lambda \log b\right) $ and $%
Z_{\psi }\left( b\right) =bZ_{T}^{-1}\left( b\right) $ that define the
relationships between fields in the original and renormalized theory. The
increase of the velocity yields a decrease of the dimensionless coupling
 $\lambda$.

Upon iterating the RG transformation one finds the RG
equations for $\lambda \left( b\right) $ as well as the temperature $T\left(
b\right)$: 
\begin{equation}
\frac{d\lambda (b)}{d\ln b}=-\lambda (b)^{2}\text{, \ }\frac{dT(b)}{d\ln b}%
=T(b)\left( 1-\lambda (b)\right) ,
\end{equation}%
These equations are solved by $T(b)=Z_{T}^{-1}\left( b\right) T$ and $%
\lambda \left( b\right) =\lambda b^{-1} Z_{T}^{-1}\left( b\right) $. The coupling 
constant is marginally irrelevant and will lead to logarithmic corrections
relative to the Dirac gas of non-interacting electrons with linear spectrum. Just
like in other quantum critical 
 phenomena, the temperature is a relevant
perturbation, causing crossover to a classical, finite $T$ regime.
Interactions lead to $T\left( b\right) <bT$,  i.e. the quantum dynamics of
the interacting Dirac liquid is more robust against thermal fluctuations
than the non-interacting Dirac gas. The coupling-constant flow equation
was obtained earlier in Ref.~\cite{Gonzalez94}.

\begin{figure}%
\epsfxsize=8.5cm
\vskip.5cm
\centerline{\epsfbox{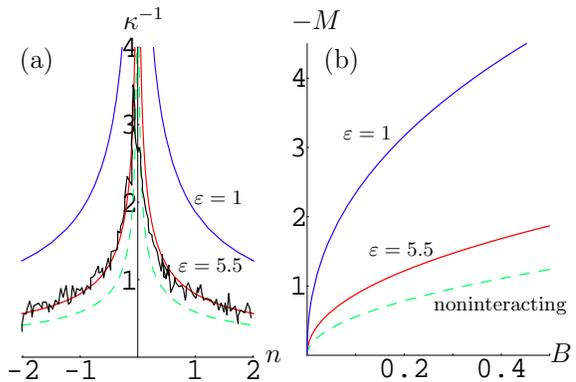}}
\vskip-.75cm
\caption{(Color Online) 
(a) Plot of the inverse compressibility $\kappa^{-1}\equiv \frac{\partial \mu}{\partial n}$ 
(in units of $(10^{-10} {\rm meV}{\rm cm}^2)$)
as a function of density $n$ (in units of $10^{12}/{\rm cm}^2$ 
for the case of 
$v = 10^6 m/s$ with the dielectric constant
 $\varepsilon = 5.5$ and $\varepsilon = 1$ (solid) 
and the noninteracting case (dashed), along
with the data from Ref.~\cite{Martin}. (b) Plot of the magnetization per area (in units of 
$\mu$A) as a function of magnetic field $(T)$. 
}
\vskip-.45cm
\label{compressfig}
\end{figure}

We now utilize the RG to develop general scaling relations for physical
observables. We start with the electron density $n$ and the
compressibility $\kappa \equiv \partial n/\partial \mu $. A finite compressibility can be
the result of a finite chemical potential $\mu $ or due to thermal
excitations at $\mu =0$. To analyze the system for finite chemical potential
we add a term $-\mu \int_{x}\psi ^{\dagger }\sigma _{0}\psi $ to the action
and rescale the fermion fields yielding $\mu \left( b\right)
=Z_{T}^{-1}\left( b\right) \mu $. Upon renormalization, 
the number of particles per area obeys:
\begin{equation}
n\left( T,\mu ,\lambda \right) =b^{-2}n\left( Z_{T}^{-1}\left( b\right)
T,Z_{T}^{-1}\left( b\right) \mu ,\lambda \left( b\right) \right) .
\label{eq:density}
\end{equation}
This equation may be written in the short-hand notation $n=b^{-2}n_{R}$, which
we shall use henceforth, with the subscript $R$ denoting renormalized quantities.  
Thus,  performing the derivative with respect to 
$\mu $ yields for the scaling of the compressibility $\kappa
=b^{-2}Z_{T}^{-1}\kappa_{R}$. As numerous similar results appear below, we
now discuss the physical meaning of this expression in some detail. The left
side of Eq.~(\ref{eq:density}) is the physical density in graphene, which
is the quantity of interest. The right side is the density in the
renormalized system where the effective coupling $\lambda (b)$ is \textit{%
small} and the effective temperature $T(b)$ is \textit{high}. With an
appropriate, physically-motivated choice for the renormalization scale $b$,
we can put the renormalized theory into a regime in which the calculation is
particularly simple. \ We fix $b$ by noting that the RG equations were
derived assuming the low-$T$ (quantum) limit, and are thus only valid for $%
T(b)<$ $T_{0}=D/\kb$ with the bandwidth $D=\hbar v\Lambda $. Thus, we choose
the renormalization condition $T(b^{\ast })=T_{0}$\cite{Millis} which yields 
$b^{\ast }=\left( 1+\lambda \log \frac{T_{0}}{T}\right) T_{0}/T$.
Approximating the renormalized high temperature compressibility by its free
fermion result, i.e. $\kappa_{R}^{-1}\simeq \pi \left( \hbar v\right)
^{2}/\left( 4 \kb T_{0}\ln 2\right) $ we obtain 
\begin{equation}
\kappa^{-1}\left( T\right) =\frac{\pi \left( \hbar v\right) ^{2}}{4
 \kb T\ln 2}\left( 1+\lambda \log \frac{T_{0}}{T}\right) ^{2},
\end{equation}
describing the nontrivial temperature dependence of the compressibility of 
 graphene, valid in the shaded region of Fig.~\ref{phasediagram}.
Since $n\left( T,\mu =0\right) =0$, we can determine the density as a function 
of $T$ for finite chemical potential $\mu \ll \kb T$ as $n(T,\lambda ,\mu
)\simeq \kappa \left( T\right) \mu $. At $T=0$ but finite $\mu $ we use $\mu
\left( b^{\ast }\right) =\hbar v\Lambda $ and it follows in full analogy 
\begin{equation}
\ n\left( \mu \right) =\frac{\mu|\mu|}{\pi \left( \hbar v\right) ^{2}\left(
1+\lambda \log \frac{D}{|\mu| }\right) ^{2}}, 
\end{equation}
and $\kappa \left( \mu \right) =\partial n/\partial \mu \simeq 2n\left( \mu
\right) /\mu $. Since it has recently been measured in scanning single
electron transistor experiments\cite{Martin} we also report the
compressibility as function of density (with $n_{0}=\Lambda ^{2}/\pi $): 
\begin{equation}
\kappa ^{-1}\left( n\right) =\hbar v\sqrt{\frac{\pi }{4|n|}}\left( 1+\frac{
\lambda }{2}\log \frac{n_{0}}{|n|}\right) , \label{compressdens}
\end{equation}
%
 that was also recently obtained by Hwang et al~\cite{Hwang0703}.

For a quantitative analysis we need to make a choice for the upper cut
off. Following essentially Ref.~\cite{Peres06} we choose $\Lambda $ such that $%
\Lambda ^{2}=2\pi /A_{0}$ where $A_{0}=3^{3/2}a_{0}^{2}/2$ is the area of
the hexagonal unit cell with C-C distance $a_{0}=1.42\times 10^{-10}\mathrm{m%
}$. It follows $T_{0}\simeq 8.34\times 10^{4}\mathrm{K}$, $D\simeq
\allowbreak 7.24\,\mathrm{eV}$ and $n_{0}\simeq $ $3.8\times
10^{15}\mathrm{cm}^{-2}$.  Thus, the arguments of the logarithms are very
large for realistic parameters of $T$, $\mu $ or $n$, making correlation
effects important.

In Fig.~\ref{compressfig}a we compare experimental data for $\kappa ^{-1}\left( n\right) $ of
Ref.~\cite{Martin} with Eq.~(\ref{compressdens}) for different dielectric
constants. For $\varepsilon =5.5$, excellent agreement between theory and
experiment is obtained. 
While the data can be fit by the Dirac gas
expression $\kappa _{0}^{-1}=\hbar v\sqrt{\frac{\pi }{4|n|}}$ with $v$ as
a fitting parameter, it cannot be understood without incorporating interactions,
since the measured velocity $v$ is known quite accurately from other experiments.
To observe the 
logarithmic variation of $\kappa$ with $n$, a larger density regime or a
combined $T$-$n$ scaling analysis would be required.

We continue our analysis with the diamagnetic susceptibility $\chi _{D}=
\frac{\partial M}{\partial B}$ of graphene, where $B$ is the magnetic field
and $M$ is the magnetization per area. It was previously shown~\cite{Nerseyan89,Ghosal} 
that, for a Dirac gas, $\chi _{D}$ diverges at $T=B=0$.
We will show that interactions enhance this divergence.  To determine the
scaling properties of $\chi _{D}$, we add a gauge-field $A_{\mu }(x)$, to
the action: $\mathcal{S}\rightarrow \mathcal{S}+\int_{x}A_{\mu }(x)j_{\mu
}\left( x\right) $ with the electrical current $j_{\mu }(x)=ev\psi ^{\dagger
}(x)\sigma _{\mu }\psi (x)$. By repeating the RG analysis in the presence of 
$A_{\mu }(x)$, we obtain the scaling properties for the Fourier transform of
the current $j_{\mu }(\mathbf{k},\omega )=Z_{J}Z_{T}^{-1}j_{\mu }(b%
\mathbf{k},Z_{T}^{-1}\omega )$ that relates the physical current to the current
in the renormalized system. The factor $Z_{J}=\lambda/\lambda(b) $ follows
from a nontrivial vertex correction for the current operator and reflects
the fact that $j_{\mu }(x)$ is a nontrivial composite operator (i.e., it
is composed of more than one $\psi $ field) that mixes high-and-low momentum
degrees of freedom.  For the diamagnetic response, we
are particularly interested in $A_{\mu }(\mathbf{r})$ that
represents a magnetic field (pointing in the $\hat{z}$ direction, normal to
the graphene sheet) via $B=\hat{z}\cdot (\mathbf{\nabla }\times \mathbf{A})$. 
Standard linear-response arguments dictate $\chi _{D}$ is related to
equilibrium fluctuations of the current:
\begin{equation}
\chi_{D}=-\mathrm{\lim }_{\bq\rightarrow 0}\frac{1}{q_xq_y}K_{xy}(
\mathbf{q},0),  \label{eq:chidresult}
\end{equation}
where $K_{\alpha \beta }(\mathbf{q},\omega )=
\left\langle j_{\alpha }(\mathbf{q},\omega )j_{\beta }(-\mathbf{q},\omega )\right\rangle$ is the current-current
correlator.  The scaling relation $K_{\alpha \beta}=Z_{T}K_{\alpha \beta ,R}$ follows directly from that of
$j_{\alpha }(\mathbf{k},\omega )$ yielding $\chi _{D}=b^{2}Z_{T}\chi _{D,R}$, and, following the scaling analysis of $\kappa$:
\begin{equation}
\chi _{D}\left( T\right) =-\frac{e^{2}v^{2}}{6\pi c^{2}\kb T}\left( 1+\lambda
\log \frac{T_{0}}{T}\right)^{2},
\label{eq:diamagresult}
\end{equation}
where we used  the diamagnetic susceptibility of non-interacting Dirac
fermions\cite{Ghosal}. 
Thus, the RG analysis has revealed a nontrivial 
\textit{enhancement\/} of the diamagnetic response (relative to the non-interacting
result) of intrinsic graphene, a result that we have 
verified perturbatively to order $\lambda$, following the procedure of Ref.~\cite{Franz}.

Our approach sheds light on the so-called universal finite-frequency
conductivity\cite{Ludwig} $\sigma (\omega )$ of graphene, related to the
current-current correlator $K_{\alpha \beta }$ via the Kubo formula $\sigma
(\omega )=\frac{1}{2\omega }\imag K_{\alpha \alpha}^{ret.}(\mathbf{q=0},\omega )$, 
where the superscript $ret.$ indicates the 
retarded function obtained via
 $i\omega
_{n}\rightarrow \omega +i0^+$. Although $K_{\alpha \beta }(\mathbf{q},\omega
)$ has  nontrivial scaling  behavior, the Kubo formula yields $\sigma
=\sigma_{R}$ for $\sigma(\omega)$. This simple scaling relation is 
constrained by a Ward identity to be valid 
to {\it all orders\/} in perturbation theory, as can be shown using an 
argument due to Gross~\cite{Gross},  and
implies that the $T\rightarrow 0$, $\omega \to 0$ (with $\omega> T$) 
universal conductivity is \textit{independent\/} of interactions. To see
this, we take (as we have above) the perturbative result for the right side,
yielding $\sigma (\omega )=\sigma (Z_{T}^{-1}\omega )=\frac{Ne^{2}}{16\hbar}$.
Thus, we find that interactions do not modify the universal conductivity, in
disagreement with recent results of Mishchenko~\cite{Mishchenko}. Again, 
this result may be verified perturbatively by computing the
leading-order corrections to the non-interacting universal conductivity;
the crucial divergent parts of the leading diagrams 
identically cancel.

Our results for the scaling properties of $\chi_{D}$, $\kappa $ etc. can be
alternatively derived from the scaling properties of the  free energy
energy density $F(T,\lambda ,\mu ,B)$. It holds $F=b^{-2}Z_{T}F_{R}$, a
result that is easy to verify to leading order in perturbation theory; more
generally it may be derived using the method of Ref.~\cite{Nelson}. The
scaling of the external magnetic field $B$ follows from the scaling of the
gauge field $\mathbf{A}$. Since within the RG we rescale positions via $%
\mathbf{r}\rightarrow \mathbf{r}^{\prime }=\mathbf{r}/b$, and $\mathbf{A}$
enters via minimal coupling, we must have for the latter $\mathbf{A}%
\rightarrow \mathbf{A}^{\prime }=b\mathbf{A}$ and therefore (from $\mathbf{B}%
=\mathbf{\nabla }\times \mathbf{A}$) $B\left( b\right) =b^{2}B$.  This
allows us  to determine the scaling of the diamagnetic moment $M=Z_{T}M_{R}$,
which yields at low $T$: 
\begin{equation}
M(B) =-\frac{3\zeta(\frac{3}{2}) }{8\pi ^{2}}\left( \frac{2e}{%
c}\right) ^{3/2}v\sqrt{B}\left( 1+\frac{\lambda }{2}\ln \frac{B_{0}}{B}%
\right),
\end{equation}
plotted in Fig.~\ref{compressfig}b.
Here $B_{0}=\hbar /\left( 2ea_{0}^{2}\right) $ is the characteristic field
where scaling stops. We used the Dirac gas expression 
\cite{Nerseyan89,Ghosal} for the magnetization at $B\left( b^{\ast }\right) =B_{0}
$. From the scaling behavior of the free energy, the scaling of $\chi _{D}$
immeadiately follows by differentiating $F(T,\lambda ,\mu ,B)$ twice with
respect to $B$, while that of $\kappa $ follows by differentiating twice
with respect to $\mu $. Similarly, one can easily derive scaling equations for
mixed derivatives such as $\partial M/\partial \mu $ or $\partial M/\partial
n$, of  interest since they  are 
 measurable for
two dimensional electron systems~\cite{Prus03}.

Turning to the heat capacity, we perform the derivative $C=-T\partial
^{2}F/\partial T^{2}$ and analyze the resulting scaling equation along the
lines of our previous calculations. It follows for the Dirac liquid:
\begin{equation}
C\left( T\right) = \frac{18T^{2}\zeta(3)}{\pi v^2\hbar^2
\left( 1+\lambda \ln \frac{T_{0}}{T}
\right) ^{2}}.  \label{heat}
\end{equation}
 The leading perturbative correction $C\left( T\right) \sim
T^{2}-2\lambda T^{2}\ln \left( T_{0}/T\right)$ fully  agrees with the recent
result of Vafek~\cite{Vafek07} who also argued that the dominant low-$T$
behavior should be $C\sim T^{2}/\ln ^{2}\frac{T_{0}}{T}$, as follows from
Eq.~(\ref{heat}).  At finite density, we obtain for the linear heat capacity
coefficient of the electron or hole Fermi liquid
$\gamma \left( n\right) \propto \sqrt{|n|}/\left( 1+\frac{\lambda }{2}\log 
\frac{n_{0}}{|n|}\right)$. 
Using Eq.~(\ref{heat}) and the fact that the spin contribution
of the magnetic susceptibility behaves as $\chi _{s}\propto c\left( T\right)
/T$ implies that the magnetic response at low $T$ is dominated by orbital
effects.

Finally, we analyze the density correlation function $\chi _{c}\left(
q\right) =\left\langle \delta n_{q}\delta n_{-q}\right\rangle $ where $%
\delta n_{q}=n_{q}-\left\langle n_{q}\right\rangle $. It determines the
dielectric function $\varepsilon \left( \mathbf{q},\omega \right) =1+\frac{%
2\pi e^{2}}{\varepsilon \left\vert \mathbf{q}\right\vert }\chi \left( 
\mathbf{q},\omega \right)$, measurable in optical or electron energy loss
scattering experiments. Due to the vanishing vertex corrections between
fermions and collective charge fluctuations we find for the density 
$n(\mathbf{k},\omega )=Z_{T}^{-1}n(b\mathbf{k},Z_{T}^{-1}\omega )$, yielding 
$\chi _{c}=Z_{T}^{-1}b^{-2}\chi _{c,R}$. If we use the random phase
approximation result for $\chi _{c,R}$  at $b^{\ast }\left\vert \mathbf{q}%
\right\vert =\Lambda $ or $Z_{T}^{-1}\left( b^{\ast }\right) \omega =D$, we
obtain $\chi _{c}\left( \mathbf{q},\omega \right) ^{-1}=2\pi e^{2}\left\vert 
\mathbf{q}\right\vert /\varepsilon -\Pi \left(\mathbf{q},\omega \right)^{-1}$,
but with the polarization (for the Dirac liquid, see also Ref~\cite{Biswas})
\begin{equation}
\Pi \left( \mathbf{q},\omega \right) =-\frac{N q^{2} }{16\sqrt{(-i\omega+0^+)^2
+v^{2}q^{2}\left( 1+\lambda \log x^{-1}\right) ^{2}}\ \ }.
\nonumber
\end{equation}
The argument of the logarithm is $x=q/\Lambda $ for $q>\frac{\omega }{\hbar
v\left( 1+\lambda \log D/\omega \right) }$ and $x=\omega /D$ otherwise. \
Despite these nontrivial renormalizations of the charge response, it follows
once more that with $\sigma \left( \omega \right) =\lim_{q\rightarrow 0}\frac{%
\omega }{q^{2}}\chi _{c}\left( q,\omega \right)$ and the microwave conductivity
of clean graphene is unaffected by interactions, as shown above.

Our results are caused by the interaction-enhanced velocity of graphene
and can be rationalized by the simple substitution $v\rightarrow v^{\ast
}=v\left( 1+\lambda \log x^{-1}\right) $ in corresponding Dirac gas
expressions, where $x$ is proportional to the dominating scaling variable,
i.e. $T$, $\sqrt{n}$, $\mu $, $\sqrt{B}$, $q$, or $\omega $. This
insight gives an immediate physical explanation for the enhanced
diamagnetism or the suppressed compressibility. The former is caused by the
increased velocity of shielding currents while the latter follows from an
increase in the density of states $\propto v^{-2}$. While the simple rule $%
v\rightarrow v^{\ast }$ is trivial to implement, we stress that it is often
the result of subtle cancellations of an infinite set of divergent diagrams
in the perturbation theory. To determine which scaling variable dominates
for a given set of parameters we use standard crossover arguments. Consider
for example temperature and density. For low density, $T\left( b\right) $
reaches its maximally allowed value $T_{0}$ before $n\left( b\right) $
reaches $n_{0}$ and $x=T/T_{0}$, the Dirac liquid regime of Fig.~\ref{phasediagram}.
 The opposite happens at higher density, giving $x=\sqrt{|n|/n_{0}}$ in the electron
and hole Fermi liquid regimes of Fig.~\ref{phasediagram}.
 The crossover from one regime to the other takes
place when $n\left( b^{\ast }\right) =n_{0}$ and $T\left( b^{\ast }\right) =T
$ simultaneously, yielding the crossover temperature $T^{\ast }\left(
n\right) $ 
%
\begin{equation}
T^{\ast}(n) =\frac{\hbar v}{\kb}\sqrt{\pi |n|}\left( 1+\frac{\lambda }{2}\log 
\frac{n_{0}}{|n|}\right) ,
\label{eq:tast}
\end{equation}
which is shown in Fig.~\ref{phasediagram}. Similar results can be obtained for any pair of
scaling variables.

In summary, by exploiting the proximity to its QCP,
we derived explicit expressions for the temperature, density and magnetic field
variation of numerous observable properties of graphene. This allows a direct comparison
with experiments to reveal the role of electron-electron correlations in this
interacting relativistic quantum liquid. 

\smallskip 
\noindent 
\textit{Acknowledgments\/} --- 
We gratefully acknowledge useful discussions with Oskar Vafek. This research
was supported by the Ames Laboratory, operated for the U.S. Department of
Energy by Iowa State University under Contract No. DE-AC02-07CH11358.

\end{document}